\documentclass[twocolumn]{aastex62}
\received{--}
\revised{--}
\accepted{--}

\shorttitle{The central star of NGC\,2346}
\shortauthors{G\'omez-Mu\~noz, M. A. et al.}

\newcommand{\teff}{$T_{\rm eff}$}
\newcommand{\lam}{$\lambda$}
\newcommand{\plus}[1]{$^{+#1}$}
\turnoffediting

\begin{document}

\title{The central star of NGC\,2346 as a clue to binary evolution through the common envelope phase}

\correspondingauthor{M. A. G\'{o}mez-Mu\~{n}oz, A. Manchado}
\email{mgomez@iac.es, amt@iac.es}

\author[0000-0002-3938-4211]{M. A. G\'{o}mez-Mu\~{n}oz}
\altaffiliation{Visiting student from the Instituto de Astrof\'isica de Canarias \\
in the Dept. of Physics and Astronomy of the Johns Hopkins \\ 
University (from September 15th 2018 to December 10th 2018).}
\affil{Instituto de Astrof\'{i}sica de Canarias, E-38205 La Laguna, Tenerife, Spain}
\affil{Departamento de Astrof\'{i}sica, Universidad de La Laguna, E-38206 La Laguna, Tenerife, Spain}
\affil{Department of Physics and Astronomy, The Johns Hopkins University, 3400 N. Charles Street, Baltimore, MD 21218, USA}

\author[0000-0002-3011-686X]{A. Manchado}
\affil{Instituto de Astrof\'{i}sica de Canarias, E-38205 La Laguna, Tenerife, Spain}
\affil{Departamento de Astrof\'{i}sica, Universidad de La Laguna, E-38206 La Laguna, Tenerife, Spain}
\affil{Consejo Superior de Investigaciones Cient{\'i}ficas, Spain}

\author[0000-0001-7746-5461]{L. Bianchi}
\affiliation{Department of Physics and Astronomy, The Johns Hopkins University, 3400 N. Charles Street, Baltimore, MD 21218, USA}

\author[0000-0002-7711-5581]{M. Manteiga}
\affiliation{Universidade da Coru\~na (UDC), Department of Nautical Sciences and Marine Engineering, 15011 A Coru\~na, Spain}
\affiliation{ CITIC, Centre for Information and Communications Technology Research, Universidade da Coru\~na, Campus de Elvi\~na sn, 15071 A Coru\~na, Spain}

\author[0000-0002-3279-9764]{R. V\'azquez}
\affiliation{Instituto de Astronom\'ia, Universidad Nacional Aut\'onoma de M\'exico, 22800 Ensenada, B.C., Mexico}



\begin{abstract}

We present an analysis of the binary central star of the planetary nebula NGC\,2346
based on archival data from the \textit{International Ultraviolet Explorer} (IUE), and new  low- and high-resolution optical spectra (3700 - 7300{\AA}).
By including in the spectral analysis the contribution of both stellar and nebular continuum, we reconciled long-time discrepant UV and optical diagnostics 
and derive $E(\bv)=0.18\pm0.01$.
We re-classified the companion star 
as A5IV by analyzing the 
wings of the Balmer absorption lines in the high-resolution ($R=67\,000$) optical spectra.
Using the distance to the nebula of 1400 pc from Gaia DR2,
we constructed a photoionization model based on abundances and line intensities
derived from the low-resolution optical spectra, and obtained a temperature of {\teff}=130\,000\,K and a 
luminosity $L=170$\,L$_\sun$ for the ionizing star,  consistent with the UV continuum. This analysis allows us to better characterize
the binary system's evolution.
We conclude that the progenitor star of NGC\,2346 has experienced a common envelope phase, in which the companion star has accreted mass and
evolved off the main-sequence. 

\end{abstract}




\section{Introduction} 

Stars with masses between $\sim$0.8--8\,M$_\sun$ end their lives as white dwarfs (WDs) after losing
most of their initial mass during the asymptotic giant branch (AGB) phase. During a brief post-AGB phase,
a planetary nebula (PN) is formed.
The simplest morphology of PNe
is well explained by the interactive stellar wind model and its generalization
where the hot core (CSPN) weak supersonic wind and radiation shape and ionize
a shell within the AGB slow wind \citep{kw78,ba87}. However, many PNe
show asymmetrical morphologies. In a 225 PNe sample, taking into account projection effects, only 20\% were found to be round and the
rest presented asymmetry \citep[63\% were elliptical and 17\% bipolar,][]{ma04}.
In fact, the fraction of bipolar PNe may be higher because the PNe would appear round if seen pole-on \citep{guma96,jomi12}.
Plausible explanations for bipolar PNe postulate a dense equatorial disk, produced by mass-loss in earlier phases, which collimates
the fast stellar wind from the hot CSPN in the post-AGB phase \citep[e.g.][]{frme94}. However, \citet{soker92}, and more
recently \citet{ga14},
have shown that a single star's angular momentum or surface rotation cannot produce sufficient equatorial density 
enhancement. AGB wind asphericities could naturally arise in a binary system,
via common envelope (CE) evolution and the initial
phase of spiraling-in \citep[][when the interaction of the companion with the red giant's atmosphere 
removes about 25\% of the CE mass]{sa98,rt12}, and gravitational focusing \citep[][in close binaries the density distribution 
of the slow wind is significantly modified by the gravity of the secondary, resulting in an enhanced density region 
close to the orbital plane of the system, and low density regions elongated perpendicularly to the orbital plane]{ga02}.
Out of more than 2000 known PNe \citep{mi12},
only 40 binary CSPN are currently known, 16 of which present orbital distances suggesting they are post-CE 
systems \citep{jo14}.
Most of them have been discovered through 
photometric variability, which favors the detection of periods shorter than 3 days, as seen in Figure 1 of \citet{dehi08}.
However, few spectroscopic binaries are known. Of those, only five CSPN have periods longer than 4\,d \citep{mi11}.

NGC\,2346 (07{$^{\rm h}$}09{$^{\rm m}$}22{\fs}52, $-$00{\arcdeg}48{\arcmin}23{\farcs}61, J2000),
is a bipolar PN with a single-lined spectroscopic binary central star (CS) with an orbital period of 16\,d
\citep[][hereafter MN81]{meni81}\defcitealias{meni81}{MN81}, recently confirmed by \citet{brown19}.
The binary system consists of the ionizing star, presumably a sdO star \citep{feal84}
with a temperature of $\sim10^5$\,K inferred by  \citet{caco78} 
using the Zanstra method,
and an A-type star companion \citep{kose73}.
\citet{feal83} obtained several IUE spectra in which the hot stellar continuum was present. 
Based
on the observed emission lines of \ion{C}{4}\,$1550\lambda$, \ion{He}{2}\,$1641\lambda$, and \ion{N}{5}$1243\lambda$, \citetalias{meni81} suggested
a {\teff} in the range of 60\,000\,K--10\,0000\,K, but they did not fit a stellar model to the spectra.
A luminosity class III of the A-type star was inferred by \citet{kose73}, using photoelectric UBV photometry,
whereas \citetalias{meni81} obtained a luminosity class V by fitting the width of the
H$\gamma$ absorption line using spectrograms. 
Later, \citet{sm97}, fitting the wings of the H$\beta$ Balmer line
in a medium-resolution spectrum, obtained a maximum value of $\log(g)=3.5$.
Recently \cite{brown19} obtained {\teff} $=7750\pm200$\,K and $\log(g)=3.0\pm 0.25$ for the cool star.

The orbital separation of this binary system, 0.16\,AU,
was calculated by \citet{mast15}, who  assumed a mass  of 1.8\,M$_\sun$ for the cool star
 with inclination angle set to that of the bipolar lobes ($i$=120\,\degr).
\citet{mast15} suggested that the system could be a remnant of CE evolution.
The common envelope is a short-lived phase in the life of a binary system during which the two stars orbit
inside a single shared envelope \citep{ivju13}. 
\citet{hato13} discussed the formation
of PNe in binary systems via a CE phase starting when a giant star overflows its Roche lobe.
Other works suggest that NGC\,2346 did not undergo CE evolution,  
mass transfer occurring instead from an evolved primary onto the companion via
Roche Lobe Overflow \citep[RLOF,][]{kori93}, or that NGC\,2346 is a remnant of
a "grazing envelope evolution" \citep{soker15}, in which the companion accreted mass
via RLOF forming an accretion disk, launching jets, and forming the lobes of
NGC\,2346.

NGC\,2346 is also remarkable because of its photometric variability, reported during certain periods of the order
of years. No luminosity variations were reported for the A5 star
until magnitude variations in form of eclipse, by up to 2 mag in the $V$ band, occurred    
from 1981 to 1986 \citep{cota86}. As we will see in section~\ref{uv_obs}, using IUE archival data, 
variations were seen until 1993.
\citet{sc85} suggested that the light variation
is related to an obscuring dust cloud  (expanding material from the hot component). Alternative explanations
have been presented by \citet{cota86}, who suggested that an ellipsoidal cool dust cloudlet with a mass of $10^{-13}$M\,$_\sun$
was responsible for the eclipse, and by \citet{peho94}, who proposed that the variation is due to the
pulsation and eclipse
of a triple system. Circumstellar dust was observed with the \textit{Spitzer} Multiband Imaging Photometer (MIPS)
\citep{su04} near the waist part of the bipolar nebula. Using \textit{Multi-conjugate adaptive optics} H$_2$ maps
(90 milliarcseconds resolution), \citet{mast15} have resolved this structure  into clumps and knots.

There are discrepancies between the $E(\bv)$ values calculated from the nebular 
H$\beta$ emission \citep[0.164--0.68,][]{alcz79,caco78,me78,phcu00} and the CS photometry \citep[0.07,][]{me78}
to NGC\,2346.
\citet{phcu00} found this discrepancy to be  strongly dependent on the extinction determination for the central star,
based mostly on photographic or photoelectric scans \citep{me78}.
Also, \citet{phcu00} found,  by means of 
narrow-band images in H$\alpha$, H$\beta$, and [\ion{O}{3}]\,{\lam}5007, that the 
extinction of the central region of the nebula is surprisingly uniform ($E(\bv)= 0.64-0.78$)
with a little evidence for reddening variation along the nebula.

We use the distance of D=1400{$\pm ^{93}_{84}$\,pc} from \citet{bj18}, derived from the Gaia DR2 parallax, for NGC\,2346.
This distance is a factor of two different from the often assumed value of $\sim$700\,pc  \citepalias{meni81} derived from a probably incorrect reddening determination.

This work aims at constraining the atmospheric parameters and the evolutionary state of the binary system of NGC\,2346,
and analyze the implications regarding the past evolution of this binary system.


\section{Observations}

\subsection{High-resolution optical spectra}

High-resolution optical spectra were taken on 2018 January 12 using the Fiber-fed Echelle Spectrograph \citep[FIES,][]{teav14} mounted on the 2.5\,m
Nordic Optical Telescope (NOT) at Roque de los Muchachos Observatory on La Palma (Spain). We used the high-resolution mode, which
provides a resolution of $R=67\,000$ in the whole visible spectral range (3700--7300\,{\AA}). The exposure time was set
to 1\,h, divided into four exposures of 900\,s, to obtain a S/N $\simeq$ 30 at 5550\,{\AA} per exposure.

FIES data were reduced with the dedicated {\sc python} reduction software FIES{\sc tool}\footnote{http://www.not.iac.es/instruments/fies/fiestool/} based on {\sc iraf}.
The standard procedures have been applied, which include bias subtraction, extraction of scattered light produced by the optical system, cosmic ray filtering, division by a normalized flat-field, wavelength
calibration by a ThAr lamp, and order merging. We combined all the merged spectra and obtained a S/N of
$\sim$62 at 5550\,{\AA}. After radial velocity correction of the spectrum,
we normalize the flux to the local continuum using iSpec \citep{blso14} by fitting a low-order polynomial to the continuum.

\subsection{Low-resolution optical spectra}

\begin{figure}
	\centering
	\includegraphics[width=0.45\textwidth]{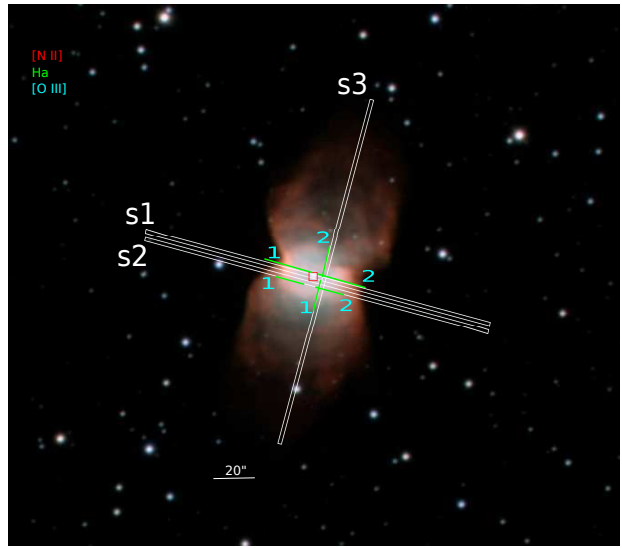}
	\caption{NGC\,2346 imaged with the 0.84\,m telescope at the OAN-SPM in three different filters.
    The picture is composed by colors: red for [\ion{N}{2}]\,\lam6584\,\AA, green for H$\alpha$, and 
    blue for [\ion{O}{3}]\,\lam5007\,{\AA}. The red square shows the position of the CS. Overplotted are the
    slit positions of the low-resolution optical spectra. North is up and East is left. \label{im:rgb_slits}}
\end{figure}

Long-slit spectra of NGC\,2346 were obtained with the Boller \& Chivens
spectrograph  mounted on the 2.1\,m telescope at the Observatorio Astron\'omico Nacional,
San Pedro M\'artir (OAN-SPM) in Mexico, during three observing runs: 2015 February 7, 9,
and 11. A E2V CCD with a 2048$\times$2048 pixel array and plate scale 
of 1.18\,{\arcsec}\,pix$^{-1}$ (in a 2$\times$2 binning mode) was used as a
detector. The 400\,lines\,mm$^{-1}$ grating was used with a 2\,{\arcsec}-wide slit yielding a 
spectral resolution of $\simeq$5.5\,{\AA} (FWHM), as judged by the arc calibration lamp spectrum, covering the 4100--7600\,{\AA} 
spectral range.
Slit positions, labeled s1--s3, are shown in Figure~\ref{im:rgb_slits}.
The position angles (PAs) for these observations were +75{\degr} for s1 and s2,
and $-$15{\degr} for s3. The exposure times were 600, 1200, and 800\,s, respectively.

The spectra were reduced using standard procedures for long-slit spectra within the {\sc IRAF}
package. We flux-calibrated the spectra by using the standard star Feige\,34.
For each slit position, the observed spectra were extracted with the \textit{APALL} task to separate 
the stellar component from the surrounding nebular emission.

Line fluxes for each extracted region were measured using the \textit{splot} task and fitting  a Gaussian function to each line.
The errors were estimated according to the RMS noise measured from flat spectral regions and then adding
$>$100  Monte Carlo simulations for each measurement. Table~\ref{tab:flux} lists the lines intensity 
(see section~\ref{optExtinction}) and the extinction coefficient ($f\lambda$) as derived from
the extinction law of \citet[][hereafter CCM89]{ccm89}. UV emission lines (Sec.\,\ref{uv_obs}) are
also included in this table.

\begin{deluxetable}{lrCC}
\tablecaption{Intrinsic line intensities of NGC\,2346. \label{tab:flux}}
\tablewidth{0pt}
\tablehead{
\colhead{Line}	&	\colhead{f\lam}	&	\colhead{s1R1\tablenotemark{a}}	&	\colhead{s1R2\tablenotemark{a}}
}
\startdata
1551$\lambda$ \ion{C}{4}       &     1.949 &     27.74$\pm$2.98    &     31.59$\pm$3.40 \\   
1640$\lambda$ \ion{He}{2}      &     1.819 &     95.34$\pm$3.10    &     108.56$\pm$3.54 \\  
1666$\lambda$ \ion{O}{3}]     &     1.796 &     6.50$\pm$1.60     &     7.41$\pm$1.82 \\  
1749$\lambda$ \ion{N}{3}]     &     1.769 &     7.98$\pm$1.69     &     9.08$\pm$1.93 \\   
1907-09$\lambda$ \ion{C}{3}]     &     1.999 &     139.80$\pm$1.90   &     159.19$\pm$2.17 \\  
3726-29$\lambda$ [\ion{O}{2}]  &     0.38  &     376.30$\pm$37.64  &     376.32$\pm$37.64 \\ 
4102$\lambda$ H$\delta$+\ion{He}{2} &0.263 &     24.97$\pm$0.33    &     25.62$\pm$0.32 \\ 
4341$\lambda$ H$\gamma$         &     0.175 &     46.60$\pm$0.29    &     46.01$\pm$0.27 \\ 
4363$\lambda$ [\ion{O}{3}]    &     0.167 &     7.98$\pm$0.16     &     7.70$\pm$0.20 \\
4471$\lambda$ \ion{He}{1}       &     0.128 &     5.54$\pm$0.25     &     6.87$\pm$0.35 \\
4686$\lambda$ \ion{He}{2}      &     0.054 &     14.73$\pm$0.27    &     16.77$\pm$0.26 \\
4711$\lambda$ \ion{He}{1}+[\ion{Ar}{4}]    &     0.046 &1.96$\pm$0.37 &     1.51$\pm$0.33 \\
4740$\lambda$ [\ion{Ar}{4}]    &     0.037 &     4.04$\pm$0.41     &     0.56$\pm$0.20 \\
4861$\lambda$ H$\beta$          &     0.0   &     100.00$\pm$0.27   &     100.00$\pm$0.30 \\
4922$\lambda$ He\,{\sc i}       &     -0.017&     2.26$\pm$0.47     &     2.01$\pm$0.30 \\
4959$\lambda$ [O\,{\sc iii}]    &     -0.027&     338.05$\pm$0.69   &     305.32$\pm$0.68 \\
5007$\lambda$ [O\,{\sc iii}]    &     -0.04 &     1002.14$\pm$1.92  &     898.54$\pm$1.93 \\ 
5198$\lambda$ [N\,{\sc i}]      &     -0.086&     7.24$\pm$0.18     &     7.74$\pm$0.28 \\ 
5518$\lambda$ [Cl\,{\sc iii}]   &     -0.149&     \nodata & 0.96$\pm$0.31 \\
5538$\lambda$ [Cl\,{\sc iii}]   &     -0.152&     0.66$\pm$0.21     &     0.50$\pm$0.24 \\   
5755$\lambda$ [N\,{\sc ii}]     &     -0.187&     7.41$\pm$0.20     &     8.12$\pm$0.19 \\    
5876$\lambda$ He\,{\sc i}       &     -0.205&     15.57$\pm$0.15    &     15.12$\pm$0.15 \\  
6300$\lambda$ [O\,{\sc i}]      &     -0.26 &     27.75$\pm$0.20    &     31.12$\pm$0.17 \\   
6312$\lambda$ [S\,{\sc iii}]+He\,{\sc ii} &-0.262 &0.59$\pm$0.09    &     0.73$\pm$0.09 \\ 
6364$\lambda$ [O\,{\sc i}]      &     -0.268&     9.07$\pm$0.22     &     9.69$\pm$0.15 \\    
6548$\lambda$ [N\,{\sc ii}]     &     -0.29 &     155.20$\pm$0.31   &     176.87$\pm$0.39 \\  
6563$\lambda$ H$\alpha$         &     -0.292&     287.94$\pm$0.57   &     287.49$\pm$0.63 \\  
6584$\lambda$ [N\,{\sc ii}]     &     -0.294&     475.51$\pm$0.93   &     535.35$\pm$1.15 \\  
6678$\lambda$ He\,{\sc i}       &     -0.305&     4.47$\pm$0.23     &     4.52$\pm$0.20 \\   
6716$\lambda$ [S\,{\sc ii}]     &     -0.31 &     7.51$\pm$0.15     &     11.25$\pm$0.16 \\   
6731$\lambda$ [S\,{\sc ii}]     &     -0.312&     6.13$\pm$0.17     &     9.97$\pm$0.16 \\    
7065$\lambda$ He\,{\sc i}       &     -0.35 &     4.03$\pm$0.17     &     4.59$\pm$0.18 \\    
7136$\lambda$ [Ar\,{\sc iii}]   &     -0.359&     27.32$\pm$0.22    &     26.74$\pm$0.19 \\   
7281$\lambda$ He\,{\sc i}       &     -0.375&     0.35$\pm$0.19     &    \nodata \\ 
7320$\lambda$ [O\,{\sc ii}]     &     -0.38 &     5.62$\pm$0.20     &     7.60$\pm$0.23 \\    
7330$\lambda$ [O\,{\sc ii}]     &     -0.381&     5.61$\pm$0.21     &     6.83$\pm$0.20 \\ \hline
$\log(F$(H$\beta$))\tablenotemark{b}					& \nodata	 &	-12.422				& -12.341	\\
\enddata
\tablenotetext{a}{All line intensities were deredened using $E(\bv)=0.18$. The intensities are with respect to $F$(H$\beta$)=100.0.
The fluxes are derived from slit 1 from region 1 and 2 (s1R1 and s2R2, respectively), as seen in Fig.\,\ref{im:rgb_slits}.}
\tablenotetext{b}{$F$(H$\beta$) in units of erg\,s$^{-1}$\,cm$^{-2}$ .}
\end{deluxetable}

\subsection{Low-resolution UV spectra}\label{uv_obs}

We  retrieved all the \textit{International Ultraviolet Explorer} (IUE) archival spectra available for NGC\,2346
from the \textit{IUE Newly Extracted Spectra} (INES)\footnote{http://sdc.cab.inta-csic.es/cgi-ines/IUEdbsMY}
and MAST\footnote{http://archive.stsci.edu/iue/}.

Short-wavelength (SW 1150--2000\,{\AA}) and long-wavelength (LW 1850--3300\,{\AA}) spectra are available, taken between 1981 to 1993, in low resolution mode (roughly 6\,{\AA}); all spectra were obtained through the large aperture (10$\times$20\arcsec). 

The spectra are listed in
Table\,\ref{tab:iue_eclipses}, with dates,  exposure times,
and orbital phase  \citepalias[calculated with $t_0$=2443142.0\,d and period of $P$ = 15.995\,d][]{meni81}.
We integrated flux in three 
 narrow continuum bands $F_{1220-1280}$ (1220--1280{\AA}), $F_{1830-1870}$ (1830--1870{\AA}), and $F_{2750-2800}$ (2750--2800{\AA})
to analyze variations among the spectra.
Fluxes were integrated avoiding bad pixels according to
the QUALITY flag, and are listed in columns 5--7 in Table\,\ref{tab:iue_eclipses}. Comments related to the quality of the spectra are also included in the last column in Table\,\ref{tab:iue_eclipses}.
UV line fluxes were measured using the \textit{splot} task in {\sc iraf} and fitting a Gaussian profile to
each line.  Errors were calculated by integrating the sigma flux (SIGMA) in the same wavelength range as the line flux. 

\begin{deluxetable*}{llCCCCCl}
\tablecaption{IUE spectra observations of NGC\,2346. \label{tab:iue_eclipses}}
\tabletypesize{\scriptsize}
\tablehead{
\colhead{Date} & \colhead{Obs. ID} & \colhead{Exp. Time} & \colhead{$\phi$\tablenotemark{a}} & 
\colhead{F$_{1288-1305}$}	& \colhead{F$_{1825-1845}$}	& \colhead{F$_{2670-2750}$} & \colhead{Comments} \\
\cline{5-7}
\colhead{} & \colhead{} & \colhead{(min)} & \colhead{} & 
\multicolumn{3}{c}{(10$^{-15}$\,erg\,s$^{-1}$\,cm$^{-2}$\,$\rm \AA^{-1}$)} & \colhead{}
}
\startdata
81/02/06	&	LWR09869		&	90	&	 0.78	&\nodata	&\nodata	&14.7\pm1.98		&	CS.\\
			&	SWP11247		&	36	&	 0.78	&\nodata	&\nodata	&\nodata			&	No spectrum visible. \\
			&	SWP11248		&	105	&	 0.78	&6.83\pm2.95	&25.67\pm1.75	&\nodata	&	CS. Geocoronal Ly$\alpha$ saturated.\\
82/02/25	&	LWR12680		&	60	&	0.76	&\nodata	&\nodata	&\nodata			&	No spectrum visible. \\
			&	SWP16420		&	50	&	0.76	&12.21\pm7.17&9.41\pm3.47	&\nodata			&	Underexposed.	\\
			&	SWP16421		&	113	&	0.77	&8.82\pm3.11	&9.78\pm1.8	&\nodata		&	CS. Very weak continuum.	\\
82/04/06	&	LWR12970		&	30	&	 0.32	&\nodata	&\nodata	&19.34\pm9.87		&	CS. Background radiation.\\
			&	SWP16704		&	40	&	 0.31	&6.08\pm21.22 &31.53\pm9.65 &\nodata	&	Underexposed. \\
82/05/05	&	LWR13172		&	60	&	 0.12	&\nodata	&\nodata	&15.61\pm5.51		&   CS. Saturated 2810--2820\AA . \\
			&	SWP16895		&	75	&	0.11  &\nodata	&\nodata	&\nodata	&	Saturated. \\
82/05/13	&	SWP16950		&	120	&	0.61	&9.21\pm3.09	&11.66\pm1.55	&\nodata			&	CS. \\
82/09/05	&	LWR14091		&	60	&	0.80	&\nodata	&\nodata	&\nodata			&	No spectrum visible. \\
			&	SWP17850		&	120	&	0.79	&11.41\pm8.08	&10.83\pm3.46 &\nodata	&	Underexposed. \ion{C}{3}] saturated.	\\
83/04/17	&	LWR15756		&	25	&	0.80	&\nodata	&\nodata	&\nodata			&	No spectrum visible. \\
			&	LWR15757		&	75	&	0.81	&\nodata	&\nodata	&\nodata			&	Saturated. \\
			&	SWP19740		&	150	&	0.80	&6.15\pm2.52	&5.68\pm1.28	&\nodata	&	Underexposed. Geocoronal Ly$\alpha$ saturated.\\
			&	SWP19741		&	105	&	0.80	&9.86\pm10.54	&13.09\pm4.0	&\nodata			&	Underexposed.\\
83/04/20	&	SWP19768		&	165	&	0.98	&10.65\pm6.52	&10.11\pm1.24	&\nodata	&	CS. Weak continuum. Geocoronal Ly$\alpha$ saturated. \\
83/05/13	&	LWR15928		&	120	&	0.42	&\nodata 	&\nodata 	&3.25\pm1.06		&	CS. \\
			&	SWP19967		&	180	&	0.42	&8.21\pm2.16	&7.68\pm1.08	&\nodata	&	CS . Geocoronal Ly$\alpha$ saturated.\\
85/02/09	&	SWP25202\tablenotemark{b}		&	415	&	0.30 &\nodata &\nodata	&\nodata 	&	Underexposed. \\
85/04/30	&	SWP25821		&	160	&	0.33	&9.52\pm3.44	&8.39\pm1.81	&\nodata	&	CS. \\
85/05/08	&	LWP05934		&	60	& 0.83	&\nodata	&\nodata	&17.08\pm2.12			&	CS. \\
			&	SWP25889		&	120 &	 0.82	&7.84\pm4.69	&26.77\pm2.14	&\nodata	&	CS. \\
86/05/03	&	SWP28258		&	150	&	0.33	&8.86\pm3.45	&30.35\pm1.59	&\nodata	&	CS. \\
86/05/07	&	SWP28266		&	120	&	0.58	&12.43\pm5.34	&34.97\pm2.13	&\nodata	&	CS. \\
93/12/15--16&	LWP27055		&	90	& 0.38	&\nodata	&\nodata	&40.34\pm1.74			&	CS. Saturated 1800--1930{\AA}. \\
			&	SWP49603		&	270	&	0.31	&10.06\pm2.47	& 37.64\pm1.17	& \nodata	&	CS. Geocoronal Ly$\alpha$ saturated. \\
\enddata
\tablenotetext{a}{Phases of the binary CSPN are based on the orbital elements of \citetalias{meni81}, t$_{0}$=2443142.0 and 15.995\,d period. The
phase shown refers to the midpoint of the exposure time.}
\tablenotetext{b}{High dispersion spectrum.}
\tablecomments{Comments are based mostly on visual inspection of the 2D images available in MAST (\url{https://archive.stsci.edu/}). 
''Underexposed'' means that there is no evidence of the CS spectrum in the 2D image, and that the spectrum has data numbers (DNs) below 100 DNs 
above background in most of the wavelength range. ``Saturated'' means that the whole spectrum in the 2D image is saturated. Spectra were taken through the large aperture (10$\times$20\arcsec).}
\end{deluxetable*}


\section{Analysis of the optical spectra}

\subsection{The A-type companion of the CSPN}

In order to derive the stellar parameters of the CSPN companion,
we compared the stellar lines of the observed high-resolution optical spectra to the library of high-resolution
solar-composition Coelho stellar models \citep{co14} by degrading the resolution of the observed
spectra to a FWHM of 0.282\,{\AA} ($R\sim20\,000$).

\begin{figure*}
	\plotone{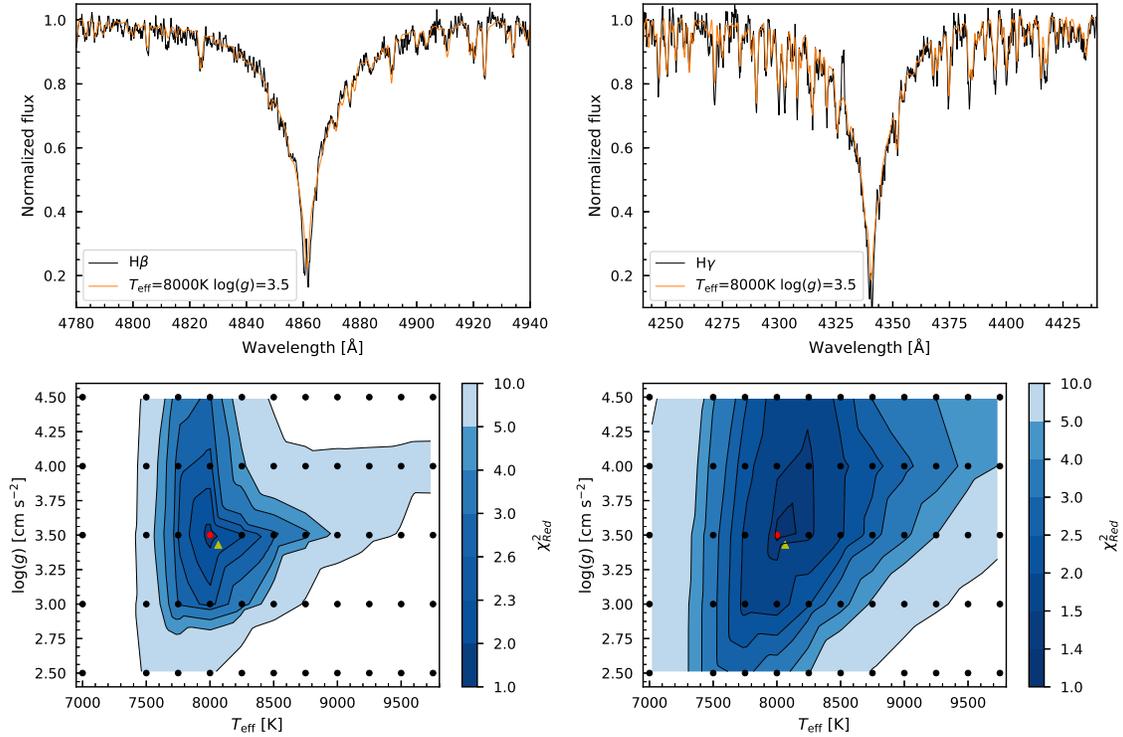}
	\caption{Mdel fit to the normalized H$\beta$ and H$\gamma$ spectral lines of NGC\,2346 (top panel).
		The best fit was obtained calculating  $\chi^{2}_{\rm Red}$ for a grid of atmospheric models (black dots); 
		fit results are represented as contour plots of different linearly-interpolated levels of $\chi^{2}_{\rm Red}$
		in the [{\teff}, log($g$)] plane.
		The best fit's $\chi^2$ is expected to be close to unity and is plotted as a red dot. The best fit result obtained
		with iSpec is also plotted (yellow triangle).
		The mean-weighted effective temperature and gravity are {\teff}=8065$\pm$180\,K and log($g$)=3.43$\pm$0.1.
		\label{im:bal_wings1}}
\end{figure*}

We then analyzed the wings of the Balmer absorption lines in the observed spectra by fitting the set of Coelho models
convolved with a FWHM of 0.282\,{\AA}.
Additionally, the models were convolved with the projected rotation velocity, $V_{\rm rot}$=47.8\,km\,s$^{-1}$,
as obtained from the stellar \ion{Mg}{2}\,{\lam}4481 absorption line by fitting a rotational profile defined by \citet{gray2005}.
We employed a reduced $\chi^2_{\rm Red}$ statistic,
\begin{equation}
	\chi^2_{\rm Red} = \frac{1}{N-k} \sum_{i=1}^{N}\left(\frac{O_i - E_i}{\sigma_i}\right)^{2}
\end{equation}

\noindent where $N$ is the number of wavelength points, $k$ is the number of free parameters (in this case just two, {\teff} and log($g$)), $E_i$ is the 
synthetic normalized spectra, $O_i$ the observed normalized spectra, and $\sigma _i$ = 1/(S/N). 
We minimized the fit for H$\gamma$ and H$\beta$ to obtain log($g$) for each one of the several plausible values of {\teff} (7000--9750\,K).
The best fit yields
{\teff}=8000$\pm$250\,K and log($g$) = 3.5$\pm$0.5 (Figure \ref{im:bal_wings1}).
The $\chi^2_{\rm Red}$ values were only estimated
in the wings of the Balmer absorption lines since the core of the lines are contaminated by the nebular emission lines. Different
values of the obtained $\chi^2_{\rm Red}$ are plotted as contours in the lower panel of Figure~\ref{im:bal_wings1}. 

We also fitted the high-resolution spectra using the spectral synthesis and modeling tool iSpec\footnote{https://www.blancocuaresma.com/s/iSpec}. A reasonable fit was
 obtained iteratively by using Kurucz model
atmospheres \citep{caku04} to produce synthetic spectra with the {\sc synthe} spectral synthesis code \citep{ku93}.
The iteration process
and $\chi^2$ minimization routine are outlined in \citet{blso14}. To obtain
the atmospheric parameters we followed the steps recommended by \citet{blso14}, varying {\teff}, log($g$), [M/H],
micro-turbulence ($v_{\rm micro}$), and macro-turbulence ($v_{\rm macro}$), and setting an initial value of $V_{\rm rot}$=2\,km\,s$^{-1}$.
The resulting
effective temperature, {\teff}=8130$\pm$130\,K, and surface gravity, log($g$)=3.43$\pm$0.10, are both consistent with our previous results.
In addition, it was possible to fit the micro-turbulence parameter, resulting in $v_{\rm micro}$=3.28\,km\,s$^{-1}$.
With all these parameters fixed, a second run with iSpec was necessary to find $V_{\rm rot}$, resulting in $V_{\rm rot}$=52$\pm$17.0\,km\,s$^{-1}$.
We have combined the values from both methods, with a statistical weight, to obtain a mean value of 
{\teff}=8065$\pm$180\,K and log($g$)=3.43$\pm$0.10,
These values, along with the $v_{\rm micro}$ value, indicate that the A-type star is
more probably a sub-giant rather than a main-sequence (MS) star \citep[][A5IV]{grgr01}.

\subsection{Extinction determination from nebular lines}
\label{optExtinction}

A reddening of $E(\bv)=0.18\pm0.01$ was obtained from a
least-squares fit to the different H$\alpha$/H$\beta$, H$\gamma$/H$\beta$,
and H$\gamma$/H$\alpha$ flux ratios for each extracted spectral region of NGC\,2346. We
assumed a Case B recombination ($n_{\rm e}=10^2$ and $T_{\rm e}=10^4$) and theoretical 
ratios of 2.863, 0.468, and 0.1635, respectively \citep{osfe06}, in conjunction
with the  extinction law of CCM89. A Monte Carlo simulation around
the flux errors was added to each line.  The results from the least-squares
fit for the different ratios were mean-weighted to obtain the final extinction coefficient value.
The errors that we are reporting are  obtained purely from
the least-squares fitting and the Monte Carlo simulation (flux errors, for the nebular Balmer lines, are less than 5\%).

The value obtained
using the Balmer decrement is in agreement with that measured by \citet{alcz79}.

\subsection{Physical conditions of the PN}

We used the  {\sc pyneb} code \citep{lumo15}, a tool for analyzing emission lines,
to calculate $n_{\rm e}$ and $T_e$ from
diagnostic diagrams using the corresponding line intensities and errors. The physical conditions were
 estimated only for the central regions in slit s1. 
To obtain the physical conditions in NGC\,2346, an extinction correction to the line intensities
of $E(\bv)=0.18$ obtained from the Balmer decrement ratio in Sect.~\ref{optExtinction} was used.
$T_{\rm e}$ was determined from [\ion{O}{3}]\,({\lam}5007+{\lam}4959)/{\lam}4363
and [\ion{N}{2}]\,({\lam}6548+{\lam}6583)/{\lam}5755, for high- and low-excitation regions, 
respectively \citep{osfe06}, and $n_{\rm e}$ was determined from [\ion{S}{2}]\,{\lam}6716/{\lam}5731.
Although the [\ion{Cl}{3}]\,{\lam}5517-37 and [\ion{Ar}{4}]\,{\lam}4711-40 emission lines are 
present in most of the regions, the uncertainty was very high and they were not used. 
The resulting electron density
$n_{\rm e}$ and temperature {$T_{\rm e}$} are reported in Table \ref{tab:phys} for the different
regions.

\subsection{Nebular abundances} \label{abund}

Ionic and elemental abundance values were obtained with the {\sc pyneb} code using 
the emission lines measured in Table \ref{tab:phys}. 
Ionization correction factors are needed because of the limited 
ionization stages observed for each element. We used the ionization correction factors (ICFs) obtained by
\citet{kiba94}. Some exceptions were the Cl and He abundances. For Cl we used the \citet{demo14} ICF because the
correction considers only the optical range, and for He we used the \citet{vaki98} ICF, which includes
the correction for collisional effects.

In order to determine the C abundance, we used the collisionally excited \ion{C}{3}]\,{\lam}1909
line from the best IUE spectrum (see Chapter\,\ref{sec:UV_analysis}),
scaled to the observed H$\beta$ flux according to the theoretical ratio of \ion{He}{2}\,1640/4686\,{\AA} (ratio of 6.474,
assuming a Case B recombination),
because no C optical recombination emission lines were found in the optical spectra. This ratio has a little temperature
and density dependence. For the O abundance, we do not
consider the [\ion{O}{2}]{\lam}7320-30 lines since they are affected by sky subtraction. We used the 
[\ion{O}{2}]\,{\lam}3726 obtained from the literature \citep{kaal76}. 

Abundances for NGC\,2346 were estimated by \citet{stgu06};  the results reported here could differ slightly
from theirs due to different apertures being used, position, and extinction correction. The elemental abundance results indicate that NGC\,2346 is a 
non-Type I PN  because of the ratio of He/H=0.1283$\pm$0.017 and N/O=0.75$\pm$0.04 \citep[Type I: He/H$\geq$0.14 and $\log($N/O$)>$0,][]{pese80}. Accordingly to 
\citet{GVD16}, these abundances indicate that the progenitor star had a mass greater than 3 M$_\sun$  and probably 
between 3.5 and 4.5 M$_\sun$, as expected from the ATON AGB models \citep{aton89} predictions and Galactic PNe sample therein.

\begin{deluxetable}{lCC}
\tablecaption{Nebular parameters. \label{tab:phys}}
\tablewidth{0pt}
\tablehead{
\colhead{Plasma Diagnostics}	&	\colhead{s1R1}	&	\colhead{s1R2}
}
\startdata
$N_{\rm e}$[\ion{S}{2}]	[cm$^{-3}$]&	179.66$\pm$43.45	&	290.23$\pm$32.22	\\
$T_{\rm e}$[\ion{N}{2}]	[K]&10146$\pm$107	&10046$\pm$88	\\
$T_{\rm e}$[\ion{O}{3}] [K]	&10647$\pm$70	&10903$\pm$92	\\
\hline
\rm Ion &   \rm s1R1    &   \rm s1R2 \\
\hline
He\plus{2} ($\times10^2$)	&11.42$\pm$0.14	&11.68$\pm$0.51	\\
He$^+$ ($\times10^2$)	&1.19$\pm$0.02	&1.35$\pm$0.02	\\
He/H ($\times10^2$)		&12.62$\pm$0.14	&13.04$\pm$0.51	\\
&	&	\\
C\plus{2} ($\times10^5$)	&21.50$\pm$2.65	&16.81$\pm$3.28	\\
C/H ($\times10^5$)		&33.11$\pm$8.32&28.18$\pm$10	\\
&	&\\
N$^+$ ($\times10^5$)	&9.22$\pm$0.17	&10.76$\pm$0.25	\\
N\plus{0} ($\times10^5$)	&0.96$\pm$0.04	&1.05$\pm$0.06	\\
N/H ($\times10^5$)		&32.00$\pm$0.95	&31.41$\pm$0.96	\\
&	&\\
O\plus{2} ($\times10^5$)	&28.86$\pm$0.09	&23.9$\pm$0.64	\\
O$^+$ ($\times10^5$)	&12.82$\pm$3.03	&13.80$\pm$3.18	\\
O\plus{0} ($\times10^5$)	&5.01$\pm$0.06	&5.71$\pm$0.20	\\
O/H ($\times10^5$)		&44.53$\pm$3.74	&40.44$\pm$3.8	\\
&	&\\
Ar\plus{3} ($\times10^6$)	&0.76$\pm$0.11	&0.22$\pm$0.06	\\
Ar\plus{2} ($\times10^6$)	&1.96$\pm$0.03	&1.82$\pm$0.04	\\
Ar/H ($\times10^6$)		&3.67$\pm$0.05	&3.42$\pm$0.07	\\
&	&\\
S\plus{2} ($\times10^6$)	&1.05$\pm$0.16	&1.16$\pm$0.15	\\
S$^+$ ($\times10^6$)	&0.34$\pm$0.01	&0.55$\pm$0.02	\\
S/H ($\times10^6$)		&1.60$\pm$0.17	&1.78$\pm$0.15	\\
&	&\\
Cl\plus{2} ($\times10^8$)	&8.02$\pm$2.48	&6.83$\pm$2.67	\\
Cl/H ($\times10^8$)		&11.42$\pm$3.41	&9.77$\pm$2.74	\\
&	&\\
N/O	&	0.71$\pm$0.03	&	0.78$\pm$0.03 \\
C/O &	0.74$\pm$0.2		&	0.70$\pm$0.25 \\
\enddata
\end{deluxetable}

\subsection{Photoionization model}
\label{phot_model}

In order to obtain the different emission components in the UV spectra 
(ionizing star + nebular continuum; Section\,\ref{uv_reddening}), we
computed a photoionization model using {\sc pycloudy} \citep{mo13}, a set of tools for dealing with the photoionization
code {\sc cloudy} v.17.00 \citep{fe17},  based on our observed emission lines and chemical abundances. The luminosity
and stellar temperature for the ionizing star, nebular diameter, and elemental abundances were optimized to
reproduce observed line fluxes in the UV/optical range, as well as that of the [\ion{O}{2}]\,3729{\lam} line from
the literature. The optimization procedure tells the code to vary one or more stellar or nebular parameters 
to reproduce the observed line intensities.

The optimized values were {\teff}=130\,000 K, $L$=170\,L$_\sun$, internal radius 0.043 parsecs, and $\log(n_{\rm H})$=2.72.

The model considers the PN as a sphere since we do not attempt to simulate an 
observation with a slit position, and the long-slit low-resolution spectra do not provide enough observational constraints to 
 model the morphology realistically. A comparison of the  observed and model line ratios is presented in 
Table~\ref{tab:model_cloudy}.

\begin{deluxetable}{llCC}
\tablecaption{Comparison of observed line ratios and those obtained with our {\sc cloudy} model. The observed flux ratios
		are the average value of s1R1 and s1R2. \label{tab:model_cloudy}}
\tablewidth{0pt}
\tablehead{
\colhead{Ion}	&	\colhead{Line}	&	\colhead{Observed}	&	\colhead{Modelled}
}
\startdata
$[$\ion{O}{3}$]$	&({\lam}5007+{\lam}4959)/{\lam}4363  &163.12	&	167.91	\\
$[$\ion{N}{2}$]$	&({\lam}6548+{\lam}6583)/{\lam}5755  &87.19	&	82.46	\\
$[$\ion{S}{2}$]$	&{\lam}6716/{\lam}6731  &1.16	&	1.14	\\ \hline
\ion{C}{4}			&{\lam}1551				&29.7	&	27.52	\\
\ion{He}{2}			&{\lam}1640				&102.0	&	112.30	\\
\ion{C}{3}$]$		&{\lam}1909				&150.0	&	159.20	\\
$[$\ion{O}{2}$]$	&{\lam}3726-29			&376.0	&	319.30	\\
H$\gamma$			&{\lam}4341				&46.1	&	47.22	\\
$[$\ion{O}{3}$]$	&{\lam}4363				&7.8	&	7.90	\\
\ion{He}{1}			&{\lam}4471				&6.2	&	6.18	\\
\ion{He}{2}			&{\lam}4686				&15.8	&	14.56	\\
\ion{He}{1}			&{\lam}4922				&2.1	&	1.66	\\
$[$\ion{O}{3}$]$	&{\lam}4959				&322.0	&	332.99	\\
$[$\ion{O}{3}$]$	&{\lam}5007				&950.3	&	993.5	\\
$[$\ion{N}{1}$]$	&{\lam}5198				&7.5	&	10.32	\\
$[$\ion{Cl}{3}$]$	&{\lam}5538				&0.5	&	0.40	\\
$[$\ion{N}{2}$]$	&{\lam}5755				&7.7	&	7.44	\\
\ion{He}{1}			&{\lam}5876				&15.3	&	15.98	\\
$[$\ion{N}{2}$]$	&{\lam}6548				&166.0	&	155.40	\\
H$\alpha$			&{\lam}6563				&287.1	&	279.35	\\
$[$\ion{N}{2}$]$	&{\lam}6583				&505.4	&	458.1	\\
\ion{He}{1}			&{\lam}6678				&4.5	&	4.46	\\
$[$\ion{S}{2}$]$	&{\lam}6716				&9.4	&	9.2		\\
$[$\ion{S}{2}$]$	&{\lam}6731				&8.1	&	8.04	\\
$[$\ion{Ar}{3}$]$	&{\lam}7136				&26.6	&	26.74	\\
\enddata
\tablenotetext{}{Note: All line intensities are with respect to $F$(H$\beta$) = 100.0.}
\end{deluxetable}

\section{Analysis of the UV spectra}
\label{sec:UV_analysis}

\subsection{Stellar parameters and UV extinction determination} \label{uv_reddening}

Our main purpose is to determine the stellar and nebular parameters, which require us to take into account
concurrently and consistently the effects of reddening. 

\begin{figure*}
\centering
\includegraphics[width=0.95\textwidth]{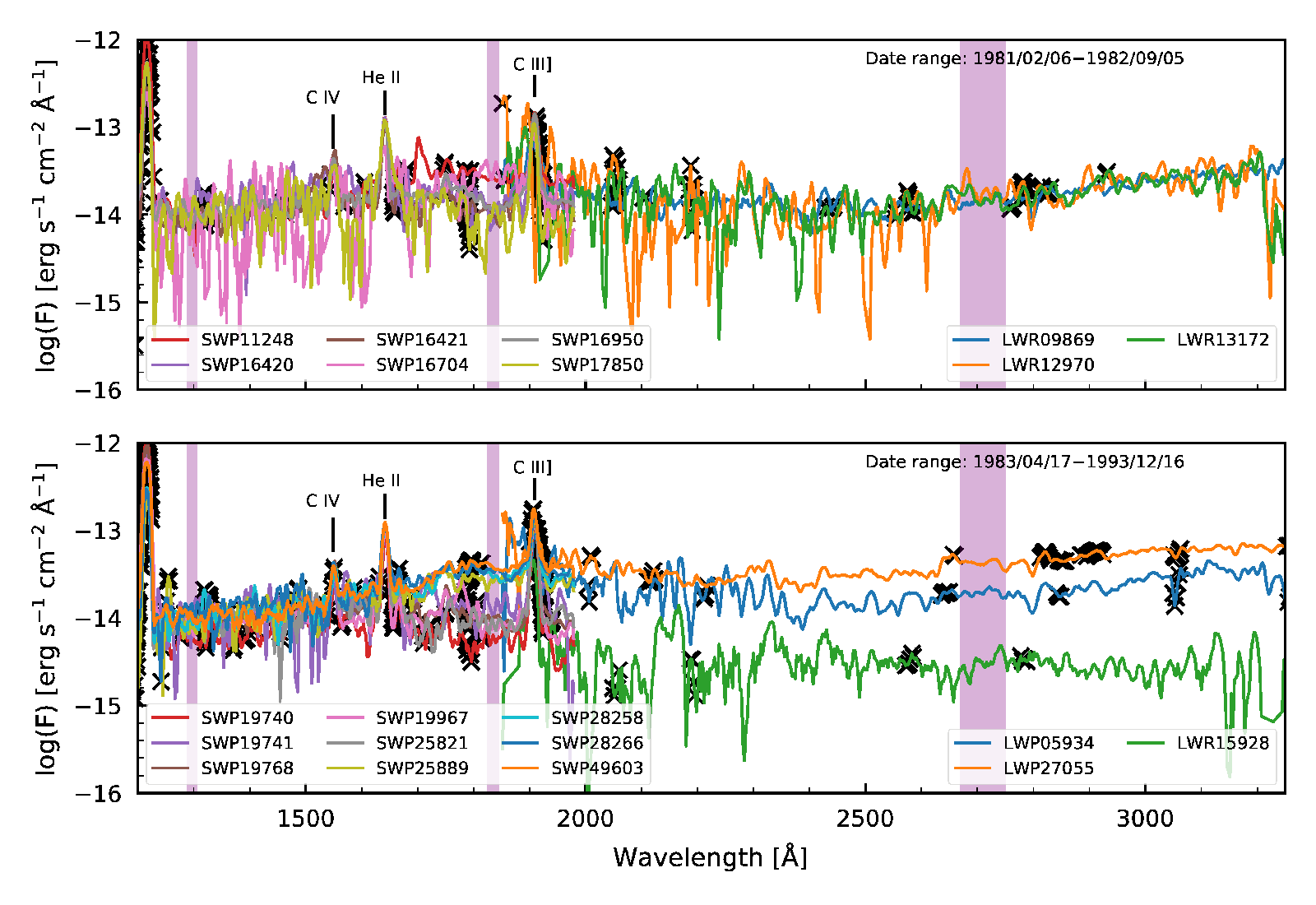}
\caption{NGC\,2346 UV IUE archival spectra. The flux around 1300\,{\AA}, that comes mainly from the hot star, is not varying. The flux around 2300\,{\AA} varies at different epochs, suggesting
a contribution from the companion star and possible eclipses. Black crosses indicate bad pixel points. The purple filled bands
indicate the width of the narrow continuum bands $F_{1288-1305}$, $F_{1825-1845}$, and $F_{2670-2750}$ (see text). \label{im:all_iue}}
\end{figure*}

In Figure~\ref{im:all_iue},
we present all the IUE spectra except for those flagged with `No spectrum visible' and `Saturated'
comments in Table~\ref{tab:iue_eclipses}.
Most of the spectra include a stellar and a nebular continuum contribution,
and nebular emission lines. All the spectra
show prominent  \ion{C}{4}, \ion{He}{2}, and \ion{C}{3}] emission lines, whose
flux varies (see Table~\ref{tab:iue_emis_lines}).
In the spectra in the upper panel of Fig.~\ref{im:all_iue}, which
span the dates between 1981/02/06--1982/09/05, the flux is almost constant, whereas in the lower panel,
the flux varies, reaching a maximum value on 1993 December 15--16 and a minimum value on 
1983 May 13. The flux variation is greater around 2800\,{\AA}
and is practically zero around 1300\.{\AA}.
The continuum
in the LWR15928 spectrum, which presents the minimum flux, shows almost exclusively nebular emission, whereas in the 
LWP27055 spectrum, which presents the maximum flux, the flux is only stellar.
We verified from the 2D spectral images that the CS was well centered in all cases.

\begin{deluxetable}{lCCC}
\tablecaption{UV emission lines in IUE spectra. \label{tab:iue_emis_lines}}
\tablehead{
\colhead{Dataset} & \colhead{\ion{C}{4}\,{\lam}1551} & \colhead{\ion{He}{2}\,{\lam}1640} &
\colhead{\ion{C}{3}]\,{\lam}1909} \\
\cline{2-4}
\colhead{(SWP)} & \multicolumn{3}{c}{(10$^{-13}$\,erg\,s$^{-1}$\,cm$^{-2}$)}
}
\startdata
11248 &	2.60\pm0.76	&	12.69\pm0.81	&14.96\pm0.48	\\
16420 &	\nodata		&	11.61\pm1.46	&15.39\pm 1.04	\\
16421 & \nodata		&	10.73\pm0.83	&17.35\pm 0.62	\\
16704 &	\nodata		&	\nodata			&8.35\pm 2.31	\\
16950 &	4.52\pm0.98	&	12.30\pm0.80	&15.86\pm 0.47	\\
17850 &	4.72\pm3.17	&	15.27\pm2.83	&12.46\pm 0.65	\\
19740 &	1.84\pm0.62	&	8.44\pm0.56		&14.18\pm 0.47	\\
19741 &	\nodata		&	7.08\pm1.76		&15.96\pm 1.18	\\
19768 &	3.60\pm0.60	&	12.07\pm0.66	&14.72\pm 0.34	\\
19967 &	2.59\pm0.54	&	11.21\pm0.87	&14.71\pm 0.32	\\
25821 &	2.74\pm1.09	&	12.73\pm0.99	&15.04\pm 0.46	\\
25889 &	2.60\pm1.35	&	12.51\pm1.29	&14.70\pm 0.67	\\
28258 &	2.51\pm1.19	&	10.10\pm0.86	&13.85\pm 0.47	\\
28266 &	3.36\pm1.60	&	11.99\pm1.61	&11.95\pm 0.55	\\
49603 &	2.01\pm0.52	&	11.17\pm0.55	&14.80\pm 0.43	\\
\enddata
\end{deluxetable}

\begin{figure}
\centering
\includegraphics[width=0.45\textwidth]{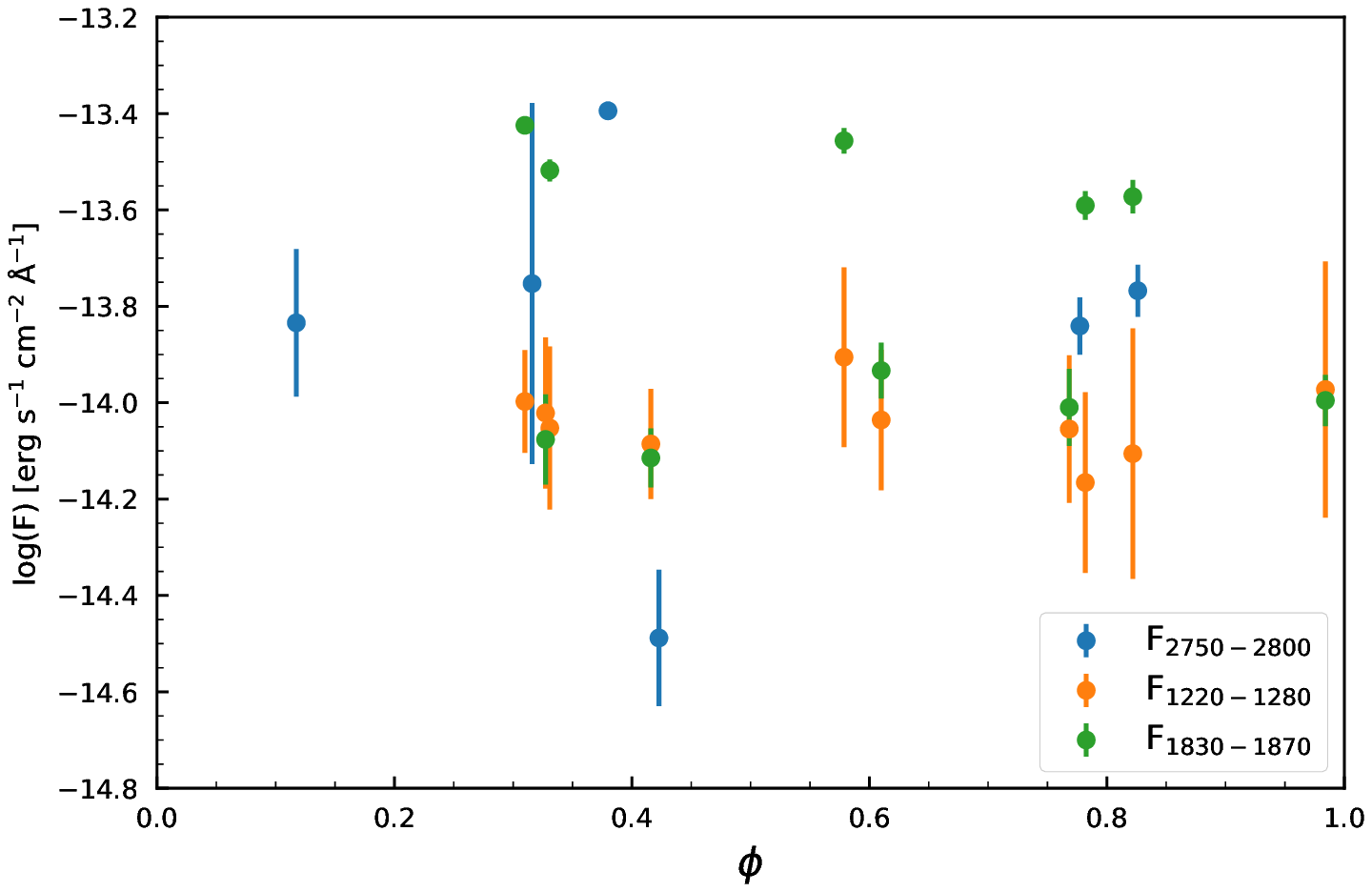}
\includegraphics[width=0.45\textwidth]{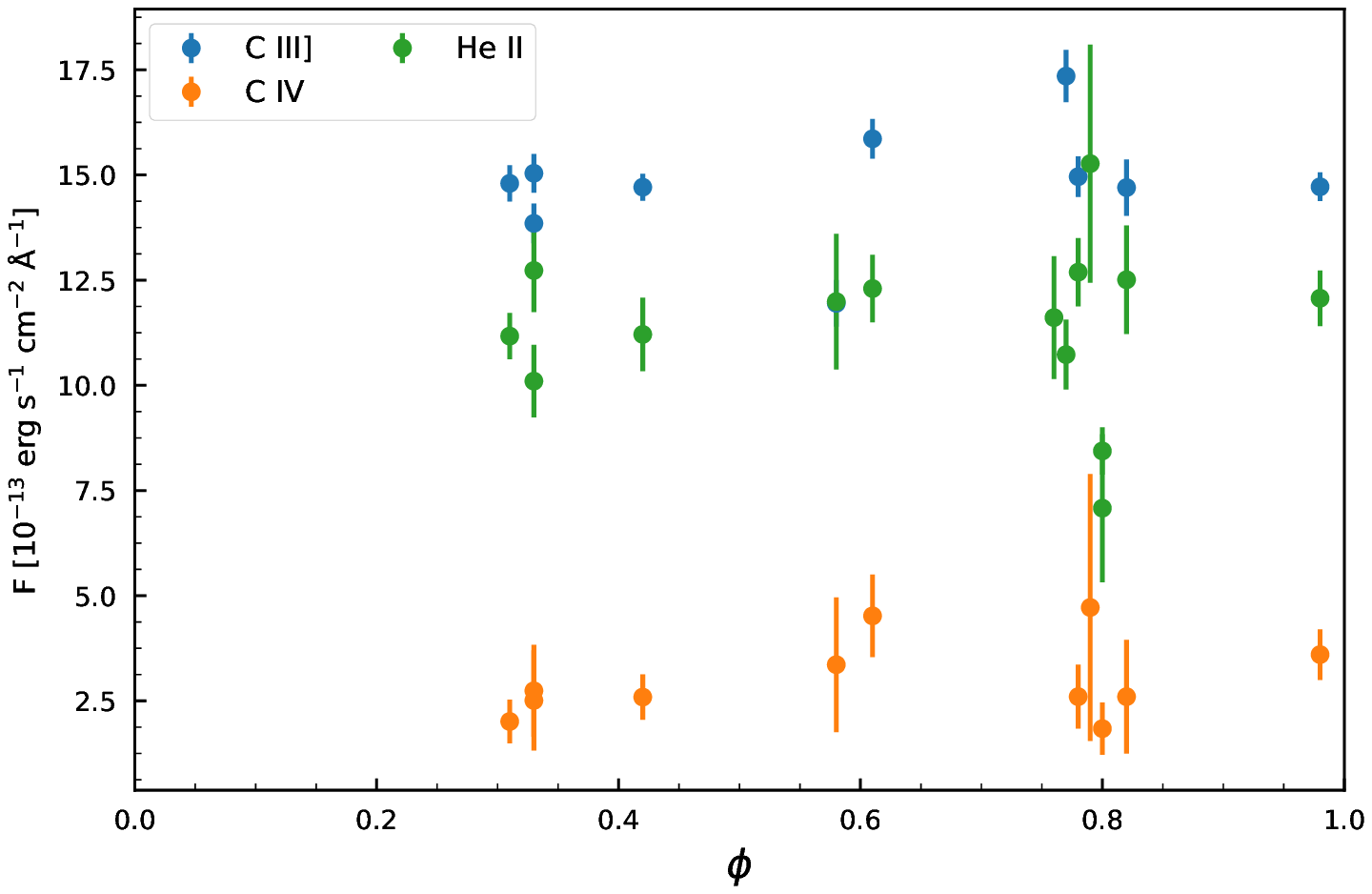}
\caption{Integrated fluxes $F_{1288-1305}$, $F_{1825-1845}$, and $F_{2670-2750}$ as a function
of the orbital phase (upper panel). $F_{1288-1305}$ is practically constant, whereas 
$F_{1825-1845}$, and $F_{2670-2750}$ show different values at different phases. The UV emission lines (lower panel)
show the same behavior as the narrow continuum bands. The variation
is probably not related to the binary orbital period. \label{im:phases_bands}}
\end{figure}

Figure~\ref{im:phases_bands} (upper panel) shows the integrated flux in narrow bands $F_{1288-1305}$, $F_{1825-1845}$, and $F_{2670-2750}$
as a function of the orbital phase ($\phi$)  for the
spectra that show stellar continuum (marked ``CS'' in Table~\ref{tab:iue_eclipses}). In the $F_{1288-1305}$ band, the flux
is practically constant. This indicates that the ionizing
source, the CSPN, is not varying, as opposed to its A-type stellar companion in the optical range \citep{cota86,peho94}.
The flux integrated in the other two continuum bands,  $F_{1825-1845}$ and $F_{2670-2750}$, varies at differing $\phi$ phases 
(refer to Table~\ref{tab:iue_eclipses} for the values).
Figure~\ref{im:phases_bands}
(lower panel) shows the emission lines flux as a function of $\phi$. It varies by a factor of $\sim$2.5 (Table~\ref{tab:iue_emis_lines}).
The flux variation of the narrow continuum bands or
the emission lines is not correlated with $\phi$.
In fact, we see a maximum brightness near
1993/12/15-16 ($\phi=0.38$) and a minimum near 1983/05/13 ($\phi=0.42$),
 the flux on 1993/12/15-16 being $\sim$13 times brighter than that on 1983/05/13 in the $F_{2670-2750}$ band (Table~\ref{tab:iue_eclipses}).
This difference is comparable
with the light variation obtained by \citet{cota86}, of  $\sim$2\,mag in the $V$-band.
It is very likely that the light variations seen in the IUE spectra are related to the 
A-type star, as suggested by \citet{feal83}.

\begin{figure}
	\centering
	\includegraphics[width=0.47\textwidth]{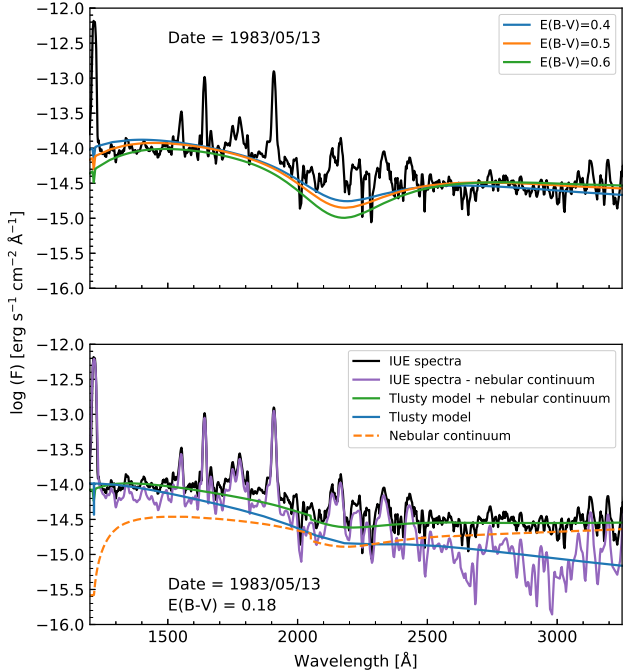}
	\caption{Upper panel: NGC\,2346 IUE archival spectra (black) and {\sc Tlusty} model with {\teff}$=$130000\,K,
	reddened with different amounts of extinction (see legend). The three reddened models have been scaled
	to match the observed flux at 2600\,{\AA}; the corresponding scaling factors imply a radius for the
	hot star between 0.071 and 0.130\,R$_\sun$, using the distance from Gaia DR2. If the IUE flux
	is the sum of a hot star and nebular continuum, we can fit the observed spectra with a lower reddening
	value (lower panel). We used the photoionization model computed in Sect.~\ref{phot_model} and 
 a reddening of  $E(\bv)=0.18$.  
	The hot star model was scaled to match the observed  1250\,{\AA} flux, since this region has no nebular continuum contribution,
	whereas the nebular continuum was scaled in such a way that the sum of the hot star plus nebular continuum matched
	the 2600\,{\AA} flux; the scaling factor implies a radius for the hot star of 0.019\,R$_\sun$.
	\label{im:iue_extinction}}
\end{figure}

For the analysis we choose IUE spectra in which the contribution of the A-type stellar companion flux is not present
and the CS continuum is prominent. As reported in  Table~\ref{tab:iue_eclipses} and seen in  Figure~\ref{im:all_iue},
the best spectra are SWP19967 and LWR15928.
We have analyzed the SWP19967 and LWR15928 spectra, fitting non-LTE plane--parallel
{\sc Tlusty} \citep{hu88} models,
which are suitable for high {\teff} and high gravity stars.
If the
IUE flux came solely from a hot stellar source, its shape would depend on stellar
{\teff} and interstellar extinction.
Using {\sc Tlusty} models, the UV continuum can be matched by a hot star model with {\teff}=125\,000\,K and  reddening values of $E(\bv)$ in the range 0.4--0.6.
The exact $E(\bv)$ value depends on the stellar model, but under the assumption that the flux is only stellar,
no acceptable fit can be found for $E(\bv)<0.4$\,mag (Fig.~\ref{im:iue_extinction}) using the extinction law of CCM89.
At a distance of 1400$\pm ^{93}_{81}$\,pc (Gaia DR2), for a model with {\teff}=125000\,K and $E(\bv)$=0.5\,mag, the match to the observed spectrum  Figure~\ref{im:iue_extinction}
(upper panel) implies a radius for the CS of $R=0.10\pm0.03$\,R$_\sun$.

We have also compared the optical magnitude for a A5V stellar companion with Kurucz models consistent with A5
types. A reddening of $E(\bv)>0.4$\,mag would imply a luminosity class for the A5 companion 
corresponding to a giant star.

However, if the IUE spectrum contains a
nebular continuum contribution, the {\sc Tlusty} model + nebular continuum will need less
reddening to match the IUE spectrum. 
To explore this possibility, the IUE spectrum was fitted with the sum
of a hot-star model and nebular continuum obtained from our photoionization model in Section~\ref{phot_model},
reddened with $E(\bv)=0.18$ (see
lower panel of Fig.~\ref{im:iue_extinction}). The hot star was scaled to match the 1250\,{\AA} flux
(since this region is not affected by nebular continuum), whereas the nebular continuum was scaled in
such a way that the sum of the hot-component plus nebular continuum matched the 2700\,{\AA} flux. The corresponding
scaling factor implies a radius for the hot star of 0.019\,R$_\sun$ (Table~\ref{tab:stellar_params_binary}),
making the scenario consistent with a post-AGB CS and the reddening consistent with the Balmer decrement.

\subsection{UV flux variations}

\begin{figure}
	\centering
	\includegraphics[width=0.45\textwidth]{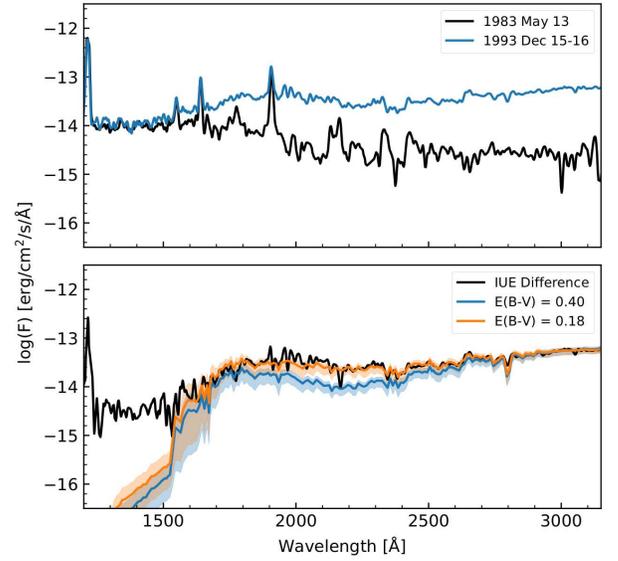}
	\caption{IUE spectra at the epochs of  maximum and
		minimum brightness observed for the A5 companion  star (upper panel). The flux difference between the two epochs (lower panel) was fitted
		using Kurucz stellar models in a  range of {\teff} and log($g$) (the solid lines represent the
		middle value and were filled between maximum and minimum values). The two extinction
		values discussed in the text are labelled in the figure. The A5 star radius,  estimated using the Gaia DR2 distance, is 4.8\,R$\sun$ or 8.8\,R$\sun$ for 
		extinction of 0.178\,mag and 0.4\,mag, respectively. \label{im:iue_two_epochs}}
\end{figure}

As  mentioned in the introduction, eclipses in the A5IV star brightness are observed in the IUE spectra (Fig.\,\ref{im:all_iue})
and in optical magnitudes \citep{ko82}.
Such eclipses are presumed to be caused by obscuring dust clouds \citep{sc85} or by an ellipsoidal cool dust
cloudlet \citep{cota86}. Since this variation does not affect the CSPN (as seen in Fig.~\ref{im:all_iue}) and the nebular continuum should not
vary (the PN is a much larger region than the orbital separation of the stars) we may infer the
stellar radius of the A5IV star by subtracting the minimum (date 1993/12/15--16) from the maximum
brightness (date 1983/05/13) detected
by the IUE (see Figure~\ref{im:iue_two_epochs} upper panel). This is
\begin{equation}
	(IS + NC + A)_{\rm max} - (IS + NC)_{\rm min} = A
\end{equation}

\noindent where $IS$ is the ionizing star brightness,
$NC$ is the nebular continuum emission,
and $A$ is the A5IV star brightness.
Scaling a solar-composition
Kurucz stellar model, using the values for {\teff} and log($g$) found from the fit of the Balmer lines,
for the Gaia distance, we obtain the stellar radius of the A5IV star.
We find a radius of $ R=4.8\pm 0.3\,R_\sun$ or $R=8.8\pm0.5\,R_\sun$, for
$E(\bv)$ of 0.18 or 0.4, respectively. The filled areas in the lower panel in Figure~\ref{im:iue_two_epochs}
represent the obtained errors in {\teff} and log($g$). However, the major uncertainty is related
to the 7\% error in the distance from which the radius errors were calculated. 

Using $E(\bv)=0.18$, derived from nebular line ratios, and the Gaia DR2 distance, values for the A5IV star 
(Table~\ref{tab:stellar_params_binary}) are: $R=4.8\pm0.3$\,R$_\sun$, 
$M=2.26\pm0.315$\,M$_\sun$ and $L=87.82\pm11.84$\,L$_\sun$. This radius for the
A5IV star suggests that it is evolved off the main-sequence.

\begin{figure*}
	\centering
	\includegraphics[width=1.05\textwidth]{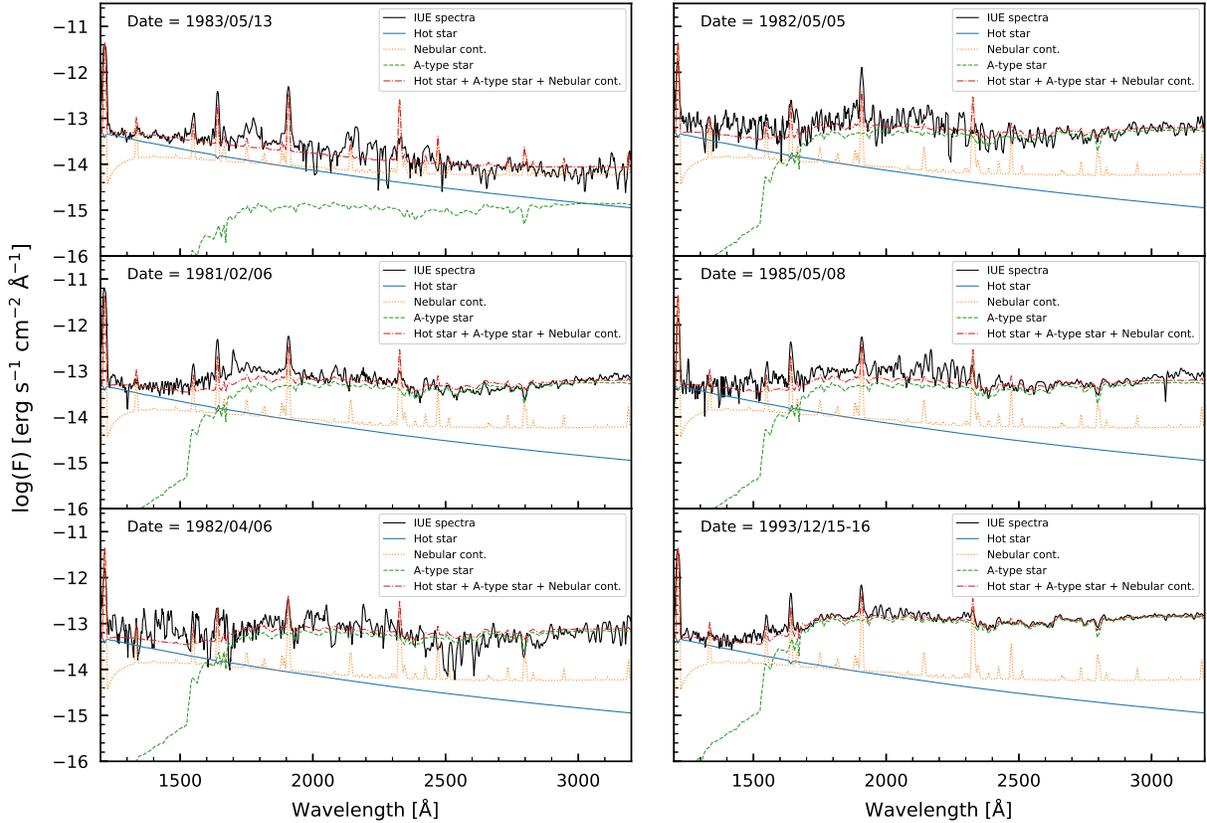}
	\caption{AIV variation of the binary CSPN of NGC\,2346 using de-reddened ($E(\bv)=0.18$) IUE spectra. The
	different flux components  and  the corresponding date of observation are labeled. There is no
	evidence of a correlation between orbital phase and eclipses of the A5IV star. \label{im:iue_eclipses}}
\end{figure*}

In Figure~\ref{im:iue_eclipses} we present the result from the IUE spectral flux decomposition into different emission components from the
photoionization model: the nebular emission,  the A5IV companion, and the CSPN for different epochs (see, Table~\ref{tab:iue_eclipses}).
The observed spectra were de-reddened
using $E(\bv)=0.18$ and the CCM89 extinction law.
The CSPN emission was fitted to match the 1200\,{\AA} region and the nebular continuum was fitted
around 2750\,{\AA} in the spectrum in which no contamination from the A5IV star was observed (spectrum 1983 May 13).
We only varied the A5IV star continuum since the eclipses do not affect the hot-star and nebular continuum emission.
Most of the spectra can be explained
with our assumptions for all components by varying the A5IV companion.
The flux variation (by 2.88\,mag) of the A5IV companion is consistent with the amplitude of the variations 
 observed by \citet{cota86} in $V$-band.

\begin{deluxetable}{lCC}
\tablecaption{Stellar parameters of the A5IV companion and the CSPN. \label{tab:stellar_params_binary}}
\tablehead{
\colhead{} & \colhead{A5IV companion} & \colhead{CSPN}
}
\startdata
{\teff}[K]	&	8065\pm180    &   130\,000	\\
$\log(g)$   &  3.43\pm0.10  &  7.0\tablenotemark{a} \\
Mass [M$_{\sun}$] &  2.26\pm0.32 & 0.7\pm_{0.3}^{0.2}  \\
Radius [R$_{\sun}$] &	4.8\pm0.3	&   0.019	\\
Luminosity [L$_{\sun}$] & 87.82\pm11.84  & 170  \\
Orbital separation [AU]      &   \multicolumn{2}{C}{0.180-0.189}   \\
$E(\bv)$[mag]   &   \multicolumn{2}{C}{0.18\pm0.01}   \\
\enddata
\tablenotetext{a}{Fixed in the photoionization model.}
\end{deluxetable}


\section{Discussion}

We estimated the extinction based on two methods: using nebular emission lines and using the UV spectrum of the CSPN. 
Our analysis of the IUE flux, accounting for CSPN continuum, nebular continuum, emission lines, and A5IV companion,
yield an upper limit for the extinction of $E(\bv)=0.18\pm0.01$.

Our results for the nebular density $n_{\rm e}\approx$200\,cm$^{-3}$ of NGC\,2346 suggest that the torus-like feature in the central region
is probably a projection effect. This is also supported by the study of \citet{mast15}, who
found that the torus-like shape is composed of clumps and cometary knots with sizes of 225--470\,AU (we rescale their results with the Gaia distance).
Such knots and clumps could also be related to the drop in brightness of the A5IV companion. 

\citet{sc85}  postulated that the variability can be explained by a cloud of material moving across the
line of sight, which was part of the shell ejected by the compact star. This is supported by
a blue-shifted component of \ion{C}{4}\,{\lam}1550 \citep{roec84}, with a radial velocity of
$\approx$1000\,km\,s$^{-1}$, which is not present in the \ion{C}{3}\,{\lam}1909 lines, indicating
that the blue-shifted emission comes from a very hot region presumably near the surface of the
CSPN. 

The temperature and luminosity of the ionizing star derived from the photoionization model are 
$L\approx170\,L_\sun$ and {\teff}$\approx130,000$\,K. Although these values cannot be taken as a unique solution
owing to the optimization process developed in {\sc cloudy} (which means that other possible solutions could satisfy
our observational constraints), such values are in good agreement with those predicted by \citet{feal84,caco78} and more
recently by \citet{mast15}, and with our analysis of the UV spectral fluxes.

\citet{mast15} calculated the mass of the ionizing star to be M$_{CS}$=0.32--0.72\,M$_\sun$ for a range of
inclination angles (120$\pm$25\,{\arcdeg} with respect to the line of sight).
Their results
were based in the assumption that the companion is an A5V star at a distance of 700\,pc.
For an A5IV secondary of mass M$_{A5}$ = 2.26$\pm$0.31\,M$_\sun$, using the Gaia DR2 distance,
the range of mass for the primary is $0.41-0.90$\,M$_\sun$. This translates into a range for the
orbital separation of 0.180--0.189\,AU ($38.70-40.64$\,R$_\sun$, see Table~\ref{tab:stellar_params_binary}).
It is very likely that the system passed through a CE phase, in which the initial orbital
separation decreased, since the ionizing star should have already reached a radius larger than
the orbital separation during its AGB (or even in the RGB) phase.
We may infer, as \citet{mast15} did, that NGC\,2346 did not result in a merger and that the minimum mass
of the PN progenitor should be greater than the mass of the A5IV companion during its main-sequence phase.	

The A5IV companion of the CSPN shares the same systemic velocity, ruling out the possibility of a 
foreground object.
 \citet{arro01} obtained a PN kinematic age of $\sim$3500\,yr (assuming a 
distance of 700\,pc) from the near-IR H$_2$ lines, whereas
\citet{es18} obtained an age of $\sim$17\,000\,yrs by using high-resolution optical spectra, from the nebular gas, and the Gaia DR1 distance
(1.285\,kpc) to model the system; the age becomes  $\sim18\,000$\,yr using the Gaia DR2 distance.
The kinematic age of the system suggests that the CSPN  should be on the constant-luminosity phase
in the H-R diagram, whereas the luminosity of the CSPN suggests that it is located on the
white dwarf cooling track, and it could be starting the final shell flash \citep{ibka83,sc85}. The discrepancy between
the kinematic age and the age of the CS could be explained by a CE phase. The CS might have  evolved through
the CE phase faster than a single star, reaching a critical effective
temperature of at least 30\,000\,K to ionize the nebula within 10\,000\,yr after the CE phase \citep{ibtu93}.
The low luminosity could also be explained by the CE scenario, since the luminosity of the remnant when
emerging from the CE phase is essentially the same as that at the start of the CE phase.
We argue that the CE phase must have started in the AGB phase of the CSPN progenitor, since this would give it  enough time to
build up a large CO core to sustain the photoionization of the nebula \citep{mast15}.
From \citet{vawo93} evolutionary tracks, the initial mass of the ionizing star, with $L$=170\,L$_\sun$ and $T_{\rm eff}$=130,000K, is close to 3.5\,M$_\sun$,
which is consistent with a remnant of $\approx$0.75\,M$_\sun$. This is also consistent with the chemical abundances derived 
in section~\ref{abund}, which indicate a mass greater than 3\,M$_\sun$, and probably between 3.5 and 4.5 M$_\sun$ for 
the progenitor star.

Comparing the stellar evolutionary time of the two stars, the CSPN progenitor would
take at least, assuming an initial mass of 3.5\,M$_\sun$, 3.3$\times10^{8}$\,yr to reach
the PN phase starting from the zero-age MS  \citep{vawo93}, whereas the companion star, at that time, would still be on the MS.
This could indicate that the system is composed
 by  a WD and a less evolved A5IV-type companion 
that has accreted mass via RLOF. This scenario can occur when the orbital separation
is less than 4\,AU \citep[e. g.,][]{soker98}, which is the case for NGC\,2346.
\citet{soker15} suggested that the secondary had launched jets that
prevented the binary system to go through a CE phase for a large fraction of the interaction time. Instead, he proposed  a new mechanism  of "grazing envelope evolution" (GEE). 
However, there is no observational evidence of jet-like features in NGC\,2346.

Not enough information exists on nebular abundances for a post-CE phase \citep[see][for a review]{de09}.
We derived ionic and elemental
abundances from our low-resolution optical spectra. From its morphology, NGC\,2346 could be considered a Type\,I PN \citep{cosc95},
although the ratio of N/O is less than 0.8. This may again indicate that the binary system
has passed through a CE phase \citep[e. g.][]{jobo15} since it is known that the binary interaction is capable of cutting
short the chemical evolution for high-mass progenitor stars \citep{de09}.

Inferences concerning the system's evolutionary stage have drastically
changed with the direct distance determination from Gaia, and our new analysis of the UV spectra taking into account the contributions of the CSPN, its companion, and nebular continuum. The analysis led to a re-classification of the A-type  companion
as an apparently sub-giant star and to constraining the stellar parameters of the CSPN. The results
suggest a binary system which evolved into a PN passing through a CE phase, 
with a A5IV companion  that has probably evolved off the main sequence due to mass accretion. The orbital period,
close to 16 days \citep{brown19}, unusually high for objects supposedly evolved in a CE scenario, provides an interesting
case study for CE evolution.


\section{Summary and Conclusions}

We have reanalyzed the binary central star of NGC\,2346 using archival IUE spectra taken at  different epochs
together with new low- and high-resolution optical spectra. We have solved the discrepancy
among extinction values reported in the literature, and derived a value of $E(\bv)=0.18\pm0.01$.
We reclassified the A5 companion, previously classified as a main-sequence star, as  A5IV, probably evolved off the main-sequence given its stellar radius (4.8\,R$_\sun$),
mass (2.26$\pm$0.315\,M{$_\sun$}), and micro-turbulence(3.28\,km\,s$^{-1}$). 
Discrepancies in the stellar evolutionary times of the CSPN and the companion star
suggest that the companion has accreted mass from the primary, increasing its mass and radius, and 
changing its evolutionary state.

We have derived {\teff}=130\,000\,K and $L$=170\,$L_\sun$ for the CSPN of NGC\,2346, from
a photoionization model based on observed nebular abundances and emission line fluxes, and by fitting the IUE spectra accounting for  a nebular continuum contribution in addition to the CSPN flux.
We propose that the CSPN of NGC\,2346
has evolved trough a CE phase, based on the inferred orbital separation and the discrepancy between evolutionary age of the CS and PN kinematic
age.

\section*{Acknowledgements}
MAGM and AM acknowledge support from the State Research Agency (AEI) of the
Spanish Ministry of Science, Innovation and Universities (MCIU) and the European
Regional Development Fund (FEDER) under grant AYA2017-88254-P. MM  acknowledges funding from
MCIU projects ESP2016-80079-C2-2-R and RTI2018-095076-B-C22 partially funded with FEDER funds,
and from Xunta de Galicia ED431B 2018/42.

This work has made use of archival data from the International Ultraviolet Exporer (IUE), obtained from the databases MAST and INES.

This article is also based on observations made with the Nordic Optical Telescope, operated by the Nordic
Optical Telescope Scientific Association at the Observatorio del Roque de los Muchachos, La Palma, Spain, of the
Instituto de Astrof\'isica de Canarias, and  on observations collected at the Observatorio Astron\'omico
Nacional at San Pedro M\'artir, B. C., Mexico.
We thank the staff at the OAN-SPM for help  obtaining our observations,
especially  Mr. Gustavo Melgoza-Kennedy and Francisco Guill\'en. We also thank the staff at the
NOT in La Palma, especially  Dr. Matteo Monelli and Dr. Joan Font.

This work has made use of data from the European Space Agency (ESA) mission
{\it Gaia} (\url{https://www.cosmos.esa.int/gaia}), processed by the {\it Gaia}
Data Processing and Analysis Consortium (DPAC,
\url{https://www.cosmos.esa.int/web/gaia/dpac/consortium}). Funding for the DPAC
has been provided by national institutions, in particular the institutions
participating in the {\it Gaia} Multilateral Agreement.

We thank the referee for helpful comments.

\software{FIESTool (\url{http://www.not.iac.es/instruments/fies/fiestool/}),
SYNTHE \citep{ku93}, ATLAS9 \citep{caku04}, IRAF \citep{tody86,tody93}, PyNeb \citep{lumo15},
CLOUDY \citep{fe17}, iSpec \citep{blso14}}




\begin{thebibliography}{99}


\bibitem[\protect\citeauthoryear{Aller \& Czyzak}{1979}]{alcz79} Aller L.~H., Czyzak S.~J., 1979, Ap\&SS, 62, 397 

\bibitem[\protect\citeauthoryear{Arias et al.}{2001}]{arro01} Arias L., Rosado M., Salas L., Cruz-Gonz{\'a}lez I., 2001, AJ, 122, 3293 

\bibitem[\protect\citeauthoryear{Bailer-Jones et al.}{2018}]{bj18} Bailer-Jones, C.~A.~L., Rybizki, J., Fouesneau, M., Mantelet, G., 
\& Andrae, R.\ 2018, \aj, 156, 58 

\bibitem[\protect\citeauthoryear{Balick, Preston, \& Icke}{1987}]{ba87} Balick B., Preston H.~L., Icke V., 1987, AJ, 94, 1641 

\bibitem[Blanco-Cuaresma et al.(2014)]{blso14} Blanco-Cuaresma, S., Soubiran, C., Heiter, U., \& Jofr{\'e}, P.\ 2014, \aap, 569, A111 

\bibitem[\protect\citeauthoryear{Brown et al.}{2019}]{brown19}Brown, A.~J., Jones, D., Boffin, H.~M.~J., \& Van Winckel, H. 2019, \mnras, 482, 4951

\bibitem[\protect\citeauthoryear{Calvet \& Cohen}{1978}]{caco78} Calvet N., Cohen M., 1978, MNRAS, 182, 687 

\bibitem[\protect\citeauthoryear{Cardelli, Clayton, \& Mathis}{1989}]{ccm89} Cardelli J.~A., Clayton G.~C., Mathis J.~S., 1989, ApJ, 345, 245

\bibitem[Castelli, \& Kurucz(2003)]{caku04} Castelli, F., \& Kurucz, R.~L.\ 2003, Modelling of Stellar Atmospheres, A20

\bibitem[\protect\citeauthoryear{Coelho}{2014}]{co14} Coelho P.~R.~T., 2014, MNRAS, 440, 1027 

\bibitem[Corradi \& Schwarz(1995)]{cosc95} Corradi, R.~L.~M., \& Schwarz, H.~E.\ 1995, \aap, 293, 871 

\bibitem[\protect\citeauthoryear{Costero et al.}{1986}]{cota86} Costero R., Tapia M., M{\'e}ndez R.~H., Echevarr{\'{\i}}a J., Roth M., Quintero A., Barral J.~F., 1986, RMxAA, 13, 149 

\bibitem[\protect\citeauthoryear{de Kool \& Ritter}{1993}]{kori93} de Kool M., Ritter H., 1993, A\&A, 267, 397 

\bibitem[De Marco et al.(2008)]{dehi08} De Marco, O., Hillwig, T.~C., \& Smith, A.~J.\ 2008, \aj, 136, 323 

\bibitem[De Marco(2009)]{de09} De Marco, O.\ 2009, \pasp, 121, 316 

\bibitem[\protect\citeauthoryear{Delgado-Inglada, Morisset, \& Stasi{\'n}ska}{2014}]{demo14} Delgado-Inglada G., Morisset C., Stasi{\'n}ska G., 2014, RMxAC, 44, 17 

\bibitem[\protect\citeauthoryear{Espinoza-Zepeda}{2018}]{es18} Espinoza-Zepeda, L.~O., 2018, BSc thesis, Universidad Aut\'onoma de Baja California

\bibitem[\protect\citeauthoryear{Feibelman \& Aller}{1983}]{feal83} Feibelman W.~A., Aller L.~H., 1983, ApJ, 270, 150 

\bibitem[\protect\citeauthoryear{Feibelman \& Aller}{1984}]{feal84} Feibelman W.~A., Aller L.~H., 1984, NASCP, 2349,  

\bibitem[\protect\citeauthoryear{Ferland et al.}{2017}]{fe17} Ferland G.~J., et al., 2017, RMxAA, 53, 385 

\bibitem[\protect\citeauthoryear{Frank \& Mellema}{1994}]{frme94} Frank A., Mellema G., 1994, A\&A, 289, 937 

\bibitem[Garc{\'{\i}}a-Hern{\'a}ndez et al.(2016)]{GVD16} Garc{\'{\i}}a-Hern{\'a}ndez, D.~A., Ventura, P., Delgado-Inglada, G., et al.\ 2016, \mnras, 461, 542 

\bibitem[\protect\citeauthoryear{Garc{\'{\i}}a-Segura et al.}{2014}]{ga14} Garc{\'{\i}}a-Segura G., Villaver E., Langer N., Yoon S.-C., Manchado A., 2014, ApJ, 783, 74 

\bibitem[\protect\citeauthoryear{Gawryszczak, Miko{\l}ajewska, \& R{\'o}{\.z}yczka}{2002}]{ga02} Gawryszczak A.~J., Miko{\l}ajewska J., R{\'o}{\.z}yczka M., 2002, A\&A, 385, 205 

\bibitem[\protect\citeauthoryear{Gray}{2005}]{gray2005} Gray D.~F., 2005, oasp.book

\bibitem[\protect\citeauthoryear{Gray, Graham, \& Hoyt}{2001}]{grgr01} Gray R.~O., Graham P.~W., Hoyt S.~R., 2001, AJ, 121, 2159 

\bibitem[Guerrero et al.(1996)]{guma96} Guerrero, M.~A., Manchado, A. \& Serra-Ricart, M.\ 1996, \apj, 456, 651.

\bibitem[Hall, et al.(2013)]{hato13} Hall, P.~D., Tout, C.~A., Izzard, R.~G., et al.\ 2013, \mnras, 435, 2048.

\bibitem[Hubeny(1988)]{hu88} Hubeny, I.\ 1988, Computer Physics Communications, 52, 103 

\bibitem[Iben et al.(1983)]{ibka83} Iben, I., Jr., Kaler, J.~B., Truran, J.~W., \& Renzini, A.\ 1983, \apj, 264, 605

\bibitem[Iben \& Tutukov(1993)]{ibtu93} Iben, I., Jr., \& Tutukov, A.~V.\ 1993, \apj, 418, 343

\bibitem[Ivanova, et al.(2013)]{ivju13} Ivanova, N., Justham, S., Chen, X., et al.\ 2013, Astronomy and Astrophysics Review, 21, 59.

\bibitem[Jones et al.(2012)]{jomi12} Jones, D., Mitchell, D.~L., Lloyd, M., et al.\ 2012, \mnras, 420, 2271.

\bibitem[\protect\citeauthoryear{Jones et al.}{2014}]{jo14} Jones D., Santander-Garcia M., Boffin H.~M.~J., Miszalski B., Corradi R.~L.~M., 2014, apn6.conf, 43 

\bibitem[Jones et al.(2015)]{jobo15} Jones, D., Boffin, H.~M.~J., Rodr{\'{\i}}guez-Gil, P., et al.\ 2015, \aap, 580, A19 

\bibitem[\protect\citeauthoryear{Kaler, Aller, \& Czyzak}{1976}]{kaal76} Kaler J.~B., Aller L.~H., Czyzak S.~J., 1976, ApJ, 203, 636 

\bibitem[\protect\citeauthoryear{Kingsburgh \& Barlow}{1994}]{kiba94} Kingsburgh R.~L., Barlow M.~J., 1994, MNRAS, 271, 257 

\bibitem[Kohoutek \& Senkbeil(1973)]{kose73} Kohoutek, L., \& Senkbeil, G.\ 1973, Liege International Astrophysical Colloquia, 18, 485 

\bibitem[\protect\citeauthoryear{Kohoutek}{1982}]{ko82} Kohoutek L., 1982, IBVS, 2113, 1 

\bibitem[Kurucz(1993)]{ku93} Kurucz, R.~L.\ 1993, Kurucz CD-ROM, Cambridge, MA: Smithsonian Astrophysical Observatory, |c1993, December 4, 1993,  

\bibitem[\protect\citeauthoryear{Kwok, Purton, \& Fitzgerald}{1978}]{kw78} Kwok S., Purton C.~R., Fitzgerald P.~M., 1978, ApJ, 219, L125 

\bibitem[\protect\citeauthoryear{Luridiana, Morisset, \& Shaw}{2015}]{lumo15} Luridiana V., Morisset C., Shaw R.~A., 2015, A\&A, 573, A42 

\bibitem[\protect\citeauthoryear{Manchado}{2004}]{ma04} Manchado A., 2004, ASPC, 313, 3 

\bibitem[\protect\citeauthoryear{Manchado et al.}{2015}]{mast15} Manchado A., Stanghellini L., Villaver E., Garc{\'{\i}}a-Segura G., Shaw R.~A., Garc{\'{\i}}a-Hern{\'a}ndez D.~A., 2015, ApJ, 808, 115 

\bibitem[Mazzitelli(1989)]{aton89} Mazzitelli, I.\ 1989, \apj, 340, 249

\bibitem[\protect\citeauthoryear{M{\'e}ndez}{1978}]{me78} M{\'e}ndez R.~H., 1978, MNRAS, 185, 647 

\bibitem[\protect\citeauthoryear{Mendez \& Niemela}{1981}]{meni81} Mendez R.~H., Niemela V.~S., 1981, ApJ, 250, 240 

\bibitem[Miszalski(2011)]{mi11} Miszalski, B.\ 2011, Monthly Notes of the Astronomical Society of South Africa, 70, 156 

\bibitem[\protect\citeauthoryear{Miszalski et al.}{2012}]{mi12} Miszalski B., Acker A., Ochsenbein F., Parker Q.~A., 2012, IAUS, 283, 442 

\bibitem[\protect\citeauthoryear{Morisset}{2013}]{mo13} Morisset C., 2013, ascl.soft, ascl:1304.020 

\bibitem[\protect\citeauthoryear{Osterbrock \& Ferland}{2006}]{osfe06} Osterbrock D.~E., Ferland G.~J., 2006, agna.book,

\bibitem[Peimbert \& Serrano(1980)]{pese80} Peimbert, M., \& Serrano, A.\ 1980, \rmxaa, 5, 9 

\bibitem[\protect\citeauthoryear{Pe{\~n}a \& Hobart}{1994}]{peho94} Pe{\~n}a J.~H., Hobart M.~A., 1994, RMxAA, 28, 165 

\bibitem[\protect\citeauthoryear{Phillips \& Cuesta}{2000}]{phcu00} Phillips J.~P., Cuesta L., 2000, AJ, 119, 335 

\bibitem[\protect\citeauthoryear{Ricker \& Taam}{2012}]{rt12} Ricker, P.~M., \& Taam, R.~E.\ 2012, \apj, 746, 74 

\bibitem[\protect\citeauthoryear{Roth et al.}{1984}]{roec84} Roth M., Echevarria J., Tapia M., Carrasco L., Costero R., Rodriguez L.~F., 1984, PASP, 96, 794 

\bibitem[\protect\citeauthoryear{Sandquist et al.}{1998}]{sa98} Sandquist, E.~L., Taam, R.~E., Chen, X., Bodenheimer, P., \& Burkert, A.\ 1998, \apj, 500, 909

\bibitem[\protect\citeauthoryear{Schaefer}{1985}]{sc85} Schaefer B.~E., 1985, ApJ, 297, 245 

\bibitem[\protect\citeauthoryear{Smalley}{1997}]{sm97} Smalley B., 1997, Obs, 117, 338 

\bibitem[Soker, \& Harpaz(1992)]{soker92} Soker, N., \& Harpaz, A.\ 1992, \pasp, 104, 923

\bibitem[Soker(1998)]{soker98} Soker, N.\ 1998, \apj, 496, 833

\bibitem[Soker(2015)]{soker15} Soker, N.\ 2015, \apj, 800, 114

\bibitem[Stanghellini et al.(2006)]{stgu06} Stanghellini, L., Guerrero, M.~A., Cunha, K., Manchado, A., \& Villaver, E.\ 2006, \apj, 651, 898  

\bibitem[\protect\citeauthoryear{Su et al.}{2004}]{su04} Su K.~Y.~L., et al., 2004, ApJS, 154, 302 

\bibitem[Telting et al.(2014)]{teav14} Telting, J.~H., Avila, G., Buchhave, L., et al.\ 2014, Astronomische Nachrichten, 335, 41 

\bibitem[Tody(1986)]{tody86} Tody, D.\ 1986, \procspie, 733

\bibitem[Tody(1993)]{tody93} Tody, D.\ 1993, Astronomical Data Analysis Software and Systems II, 173

\bibitem[Vassiliadis \& Wood(1993)]{vawo93} Vassiliadis, E., \& Wood, P.~R.\ 1993, \apj, 413, 641 

\bibitem[\protect\citeauthoryear{V\'{a}zquez, Kingsburgh, \& L\'{o}pez}{1998}]{vaki98} V\'{a}zquez R., Kingsburgh R.~L., L\'{o}pez J.~A., 1998, MNRAS, 296, 564 

\end{thebibliography}
\end{document}